\documentclass[journal=jacsat,manuscript=article,layout=twocolumn]{achemso}
\usepackage[version=3]{mhchem} 

\usepackage[utf8]{inputenc}
\usepackage{graphicx}
\usepackage{amsmath}
\usepackage{float}
\usepackage{amssymb}
\usepackage{placeins}
\usepackage{array}
\usepackage{color}
\usepackage{physics}
\usepackage{makecell}
\usepackage{siunitx}
\usepackage{xstring}
\usepackage{multirow}
\usepackage{mathtools}
\usepackage{soul} 
\usepackage{changes}
\usepackage{dcolumn}
\usepackage{bm}
\usepackage{subcaption} 
\usepackage{makecell}
\usepackage{soul}
\usepackage{comment}
\usepackage{hyperref} 
\hypersetup{
    colorlinks=true,
    linkcolor=red,
    filecolor=cyan,      
    urlcolor=blue,
    citecolor=blue,
    bookmarks=true,
    pdfpagemode=FullScreen
}

\DeclareUnicodeCharacter{0301}{\'{e}}
\DeclareUnicodeCharacter{0301}{\'{y}}
\DeclareUnicodeCharacter{0304}{\'{i}}
\DeclareUnicodeCharacter{0308}{\"{o}}

\newcommand{\rn}[1]{%
  \textup{\uppercase\expandafter{\romannumeral#1}}%
}

\newcommand{\rates}{{\tau^{-1}_s}}

\newcommand{\cspbbr}{\mathrm{CsPbBr_3}}
\newcommand{\mapbbr}{\mathrm{MAPbBr_3}}
\newcommand{\mpsnbr}{\mathrm{MPSnBr_3}}
\def\frohlich{\text{Fr{\"o}hlich}}
\usepackage{caption}

\definecolor{kc}{rgb}{0.6,0.2,1}

\author{Kejun Li}
\affiliation[University of California, Santa Cruz]
{Department of Physics, University of California, Santa Cruz, California, 95064, United States}
\alsoaffiliation[University of Wisconsin-Madison]
{Department of Materials Science and Engineering, University of Wisconsin-Madison, 53706, United States}
\author{Junqing Xu}
\affiliation[Hefei University of Technology]
{Department of Physics, Hefei University of Technology, Hefei, Anhui, 230009, China} 
\author{Uyen N. Huynh}
\affiliation[University of Utah]
{Department of Physics and Astronomy, University of Utah, Salt Lake City, UT, 84112, United States} 
\author{Rikard Bodin}
\affiliation[University of Utah]
{Department of Physics and Astronomy, University of Utah, Salt Lake City, UT, 84112, United States} 
\author{Mayank Gupta} 
\affiliation[University of Wisconsin-Madison]
{Department of Materials Science and Engineering, University of Wisconsin-Madison, 53706, United States}
\author{Christian Multunas}
\affiliation[Rensselaer Polytechnic Institute]
{Department of Physics, Rensselaer Polytechnic Institute, 110 8th Street, Troy, New York 12180, United States}  
\author{Jacopo Simoni}
\affiliation[University of Wisconsin-Madison]
{Department of Materials Science and Engineering, University of Wisconsin-Madison, 53706, United States}
\author{Ravishankar Sundararaman}
\affiliation[Rensselaer Polytechnic Institute]
{Department of Materials Science and Engineering, Rensselaer Polytechnic Institute, 110 8th Street, Troy, New York 12180, United States}  
\author{Zeev Valy Verdany} 
\affiliation[University of Utah]
{Department of Physics and Astronomy, University of Utah, Salt Lake City, UT, 84112, United States}  
\author{Yuan Ping}
\email{yping3@wisc.edu}
\affiliation[University of Wisconsin-Madison]
{Department of Materials Science and Engineering, University of Wisconsin-Madison, 53706, United States}
\alsoaffiliation[University of Wisconsin-Madison]
{Department of Physics, University of Wisconsin-Madison, Madison, Wisconsin 53706, United States}
\alsoaffiliation[University of Wisconsin-Madison]
{Department of Chemistry, University of Wisconsin-Madison, Madison, Wisconsin 53706, United States}

\title{Spin Dynamics in Hybrid Halide Perovskites  - Effect of Dynamical and Permanent Symmetry Breaking}

\begin{document}
\begin{abstract}
The hybrid organic-inorganic halide perovskite (HOIP), for example $\mapbbr$, exhibits extended spin lifetime and apparent spin lifetime anisotropy in experiments. The underlying mechanisms of these phenomena remain illusive. By utilizing our first-principles density-matrix dynamics approach with quantum scatterings including electron-phonon and electron-electron interactions and self-consistent spin-orbit coupling, we present temperature- and magnetic field-dependent spin lifetimes in hybrid perovskites, in agreement with experimental observations. For centrosymmetric hybrid perovskite MAPbBr$_3$, the experimentally observed spin lifetime anisotropy is mainly attributed to the dynamical Rashba effect arising from the interaction between organic and inorganic components and the rotation of the organic cation. For noncentrosymmetric perovskite, such as $\mpsnbr$, we found persistent spin helix texture at the conduction band minimum, which significantly enhances the spin lifetime anisotropy. Our study provides theoretical insight to spin dynamics in HOIP and strategies for controlling and optimizing spin transport.
\end{abstract}


\section{Introduction}
The field of semiconductor spintronics has gained significant attention in recent years, due to its applications in low-power electronics~\cite{privitera2021perspectives,sierra2021van,he2022topological,kim2022ferrimagnetic,yang2021chiral}. Spintronics utilizes spin as the information carrier, offering a promising alternative to traditional charge-based electronics. However, one main challenge in creating efficient spintronic devices is to achieve both long spin lifetimes and strong spin-orbit coupling (SOC) simultaneously~\cite{lu2024spintronic,bader2010spintronics,vzutic2004spintronics}. These two properties are essential for retaining stable information and enabling the precise control of spin states.

Halide perovskites have emerged as strong candidates for spintronic applications due to their unique combination of long spin lifetimes, strong SOC, and tunable symmetry~\cite{kim2021chiral,belykh2019coherent,huynh2022transient,zhou2020effect,zhang2022room,Ping2018}. Some of these materials also exhibit large Rashba splitting\cite{chen2021tunable,lu2024spintronic} and persistent spin helix (PSH) texture with SU(2) symmetry~\cite{zhang2022room}, which highlight their potential for future spintronics applications.

In this first-principles study, we investigate the spin relaxation and dephasing in the three-dimensional hybrid perovskite systems under zero and finite magnetic fields. We found that the spin relaxation of centrosymmetric $\mapbbr$ is nearly isotropic and closely resembles that of $\cspbbr$~\cite{xu2024spin,huynh2022transient},
when both materials have the same $Pnma$ crystal symmetry. The spin dephasing under magnetic field is slightly anisotropic due to the weak anisotropy of  $g$ factor fluctuation. Therefore, this alone does not explain the large anisotropy of spin lifetime (up to 2.5) observed experimentally. Interestingly, when the symmetry is broken due to structural dynamics such as molecular rotation (which has a similar time scale with spin lifetime at room temperature), the spin relaxation becomes anisotropic, attributed to the spin split induced by symmetry broken. This trend and spin lifetime anisotropy ratio align with the experimental observation, and underscores the importance of structural dynamics and symmetry on spin relaxation.

Finally, we show that $\mpsnbr$, with a permanently broken symmetry, exhibits significant spin lifetime anisotropy in the range of 20-50. In particular, we find that this large anisotropy correlates with its PSH spin texture, and the corresponding spin lifetime parallel to spin texture shows dominantly spin-flip Elliot-Yafet mechanism. Our work points out the strong tunability of spin dynamics properties in halide pervoskites, and provides fundamental insights to guide materials design of pervoskites for opto-spintronics applications. 


\section{Results and Discussion}

\subsection{Spin Lifetime of $\mapbbr$ with Centro-symmetry}
By employing our first-principles density-matrix dynamics (FPDM) method (more details in computational detail), first we focus on the intrinsic spin relaxation of centrosymmetric $\mapbbr$ at $Pnma$ crystal symmetry. This is crucial for understanding its behavior under more complex external conditions. 
For the intrinsic spin relaxation, we include both electron-electron (e-e) and electron-phonon (e-ph) scattering processes. Since the e-e scattering process is negligible compared to electron-phonon in this system (shown in SM Fig.~\textcolor{red}{S1}), its spin relaxation is mainly attributed to electron-phonon scatterings and spin-orbit couplings. 

\begin{figure*}
    \centering
    \includegraphics[width=\textwidth]{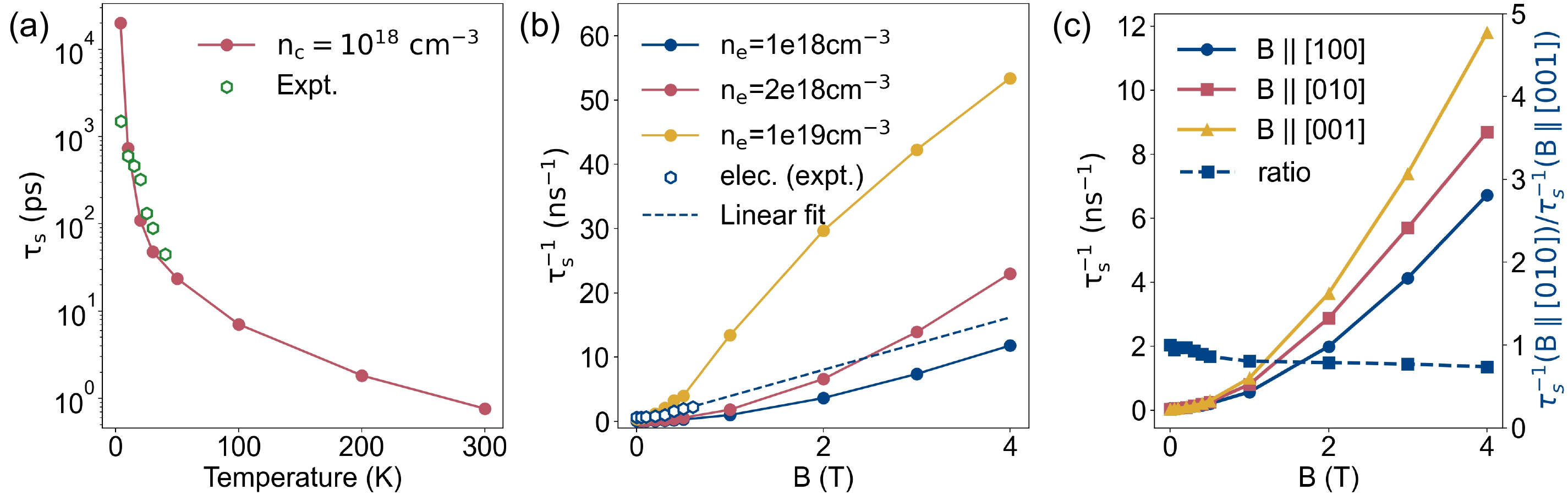}
    \captionsetup{justification=justified,singlelinecheck=false}
    \caption{Electron Spin lifetime of centrosymmetric $\mapbbr$. (a) Temperature-dependent electron spin lifetimes $\tau_{s,z}$ from the FPDM calculations and experiments~\cite{huynh2024magneto}. 
    The experiment is performed at a laser power of 800 $\si{\mu W}$, corresponding to carrier concentration of $10^{18}$ $\si{cm^{-3}}$. 
    (b) Magnetic field-dependent spin dephasing rate of electrons at T = 4 K,  with carrier concentrations of $1\times10^{18}$, $2\times10^{18}$ and $1\times10^{19}$ $\si{cm^{-3}}$. The experimental data measured at $P=100$ mW, with the dashed line as a linear fit of last three points. $\tau_{s,x}$ is evaluated with the external $\mathbf{B}$ along [001]. 
    (c) Anisotropy of spin dephasing rate of electrons. $\tau_{s,y}$ is evaluated for $\mathbf{B}\parallel[100]$, and $\tau_{s,x}$ is evaluated for $\mathbf{B}\parallel[010]$ and $\mathbf{B}\parallel[001]$.  
    The data points with the blue dashed line represent the ratio of $\rates(B\parallel [010])/\rates(B\parallel [001])$.
    }
    \label{fig:tau_s_vs_T_nc}
\end{figure*}

As shown in Fig.~\ref{fig:tau_s_vs_T_nc}(a), under zero magnetic field, the spin relaxation time ($T_1$) in $\mapbbr$ decreases with increasing temperature. This trend is primarily caused by an increased phonon occupation at higher temperatures, and correspondingly, stronger electron-phonon scattering strength, facilitating spin-flip transition dominated by the Elliot-Yafet (EY) mechanism~\cite{vzutic2004spintronics,fabian1998spin}, as can be seen in SM Fig.~\textcolor{red}{S3}. Such temperature dependence is similar to what we found in CsPbBr$_3$ earlier~\cite{xu2024spin}.

The phonon contribution to electron-phonon interaction in $\mapbbr$ involves complex molecular-inorganic lattice hybrid vibrational modes due to the inclusion of the organic MA molecule comparing with previous studies on $\cspbbr$~\cite{xu2024spin}.
Contrary to the common assumption that LO phonons via the $\frohlich$ interaction are the major contributors to spin relaxation~\cite{belykh2019coherent,crane2020coherent,sio2022unified}, the spin relaxation is primarily caused by lower-energy optical phonons 
in the energy range from 2-6 meV as shown by the phonon dispersion in SM Fig.~\textcolor{red}{S4}. This contribution of phonon is evidenced by the spin lifetimes determined by electron interacting with different groups of phonons in SM Fig.~\textcolor{red}{S5}(a). In particular, 
the spin-flip electron-phonon matrix elements in SM Fig.~\textcolor{red}{S6}(a),  which are the most relevant to spin relaxation, show distinct phonon dependence from the spin-conserving matrix elements. For example, the 2-6 meV optical phonon modes as the major contributors to spin relaxation,  have negligible contributions to the spin-conserving matrix elements.
The different phonon contribution to
spin relaxation can be attributed to the short-range nature of spin-phonon interaction as spin indirectly interacts
with phonon via SOC.
On the other hand, the $\frohlich$ like phonon modes, giving rise to a large LO-TO splitting at $13-21$ meV, consistent with the strong Raman intensity at 12.3 meV, 12.6 meV, 16.2 meV and $22.3\pm0.6$ meV~\cite{matsuishi2004optical, ledinsky2015raman}. 
They have large contribution to carrier relaxation through spin-conserving electron-phonon scattering processes, little to spin-flip processes or spin relaxation, as shown in SM Fig.~\textcolor{red}{S5}(b) and SM Fig.~\textcolor{red}{S6}(b). 

Next, we look at the spin dynamics of $\mapbbr$ at finite magnetic field. The $k$-dependent spin vectors, 
can undergo inhomogenous precession in an external magnetic field that is perpendicular to its initial spin direction, leading to spin dephasing, fundamentally because of $g$ factor induced magnetic field fluctuations~\cite{vzutic2004spintronics}. Specifically, the interaction of spin with an external magnetic field $\mathrm{{\bf B}}$ is described by the following Hamiltonian approximately,
\begin{align}
H_{k}\left({\bf B}\right)= H_{0,k}+\mu_{B}{\bf B}\cdot\left({\bf L}_{k}+g_{0}{\bf S}_{k}\right)
\end{align}
where $\mu_{B}$ is the Bohr magneton, $g_{0}$ is the free-electron $g$-factor, and ${\bf S}$ and ${\bf L}$ are the spin and orbital angular momentum, respectively. We calculated ${\bf L}$ using Berry phase formalism with details in  Refs.~\cite{multunas2023circular,xu2024spin}. Under the external magnetic field, the total spin decay rate under a magnetic field can be expressed as,
\begin{align}
\tau_{s}^{-1}\left(\mathrm{{\bf B}}\right) \approx \left(\tau_{s}^{\mathrm{0}}\right)^{-1} + \left(\tau_{s}^{\Delta\Omega}\right)^{-1}\left(\mathrm{{\bf B}}\right)~\label{eq:tot_spin_decay}
\end{align}
where $(\tau_{s}^{\mathrm{0}})^{-1}$ is the zero-field spin relaxation rate, $(\tau_{s}^{\Delta\Omega})^{-1}$ is the spin dephasing rate under external B field, and $\Delta\Omega$ is 
the fluctuation of the Larmor-precession-frequency expressed as,
\begin{align}
     \Delta\Omega &= \mu_{B} B \Delta \widetilde{g}
\end{align}
where $\Delta \widetilde{g}$ is the fluctuation (or variation) of $g$ factor, whose detailed derivation can be found in Ref.~\cite{xu2024spin}, with results shown in SM Fig.~\textcolor{red}{S7}. 
$(\tau_{s}^{\Delta\Omega})^{-1}$ spin dephasing rate has been mainly discussed with Dyakonov-Perel (DP) mechanism at strong scattering limit ($\tau_{p}\Delta\Omega \ll 1$), 
 or free induction decay (FID) mechanisms at weak scattering limit ($\tau_{p}\Delta\Omega \gtrsim 1$)~\cite{wu2010spin}. 
 

In Fig.~\ref{fig:tau_s_vs_T_nc}(b), we show the spin dephasing at 4 K, at which temperature the spin relaxation due to electron-phonon coupling is weak, where the experimental reference in this work was carried out~\cite{huynh2024magneto}. In comparison to the experiment, which is carried out at $P=100$ mW, corresponding to the carrier density $n_c = 10^{18}~\si{cm^{-3}}$, our theoretical results show good agreement when $B^\text{ext} < 0.5$ T. The experimental data locate in-between $n_c= 1\times10^{18}~\si{cm^{-3}}$ to $1\times10^{19}~\si{cm^{-3}}$ of the calculated results. The onset of the linear dependence on the magnetic field (FID mechanism) occurs earlier in experiments than theory. Additionally, a slightly higher spin dephasing rate has been observed experimentally, consistent with previous work on $\cspbbr$~\cite{xu2024spin}. This discrepancy may be attributed to the presence of nuclear spins. By estimation, the hyperfine interaction is mainly from the electron spin interacting with the nuclear spin of Pb and Br atoms, which ranges from $4\times10^{8}$ Hz to $6\times10^{9}$ Hz. The spin dephasing time ($T_2^*$) of localized carriers due to the hyperfine interaction is 0.6–8.0 ns for electrons and 0.35-4.60 ns for holes, setting the upper bound of spin lifetime at $B=0$ T~\cite{xu2024spin}. In Fig.~\ref{fig:tau_s_vs_T_nc}(a), the discrepancy between theory and experiment at T$<10$ K could also be attributed to the hyperfine interaction-induced spin relaxation. 

An interesting fact from our calculation is that the spin dephasing rate shows slight anisotropy between [001] and [010] at a finite magnetic field, with the factor of rate anisotropy ($\rates(B\parallel [010])/\rates(B\parallel [001])$) about 0.8,  
as can be seen in Fig.~\ref{fig:tau_s_vs_T_nc}(c). 
However, the experimental results~\cite{huynh2024magneto} show a factor of anisotropy ${\sim}2.5$ between $B\parallel[010]$ and $B\parallel[001]$. 
The stronger spin relaxation anisotropy in experiments indicates that some additional mechanisms may mediate spin relaxation and dephasing as discussed below.

\subsection{Strong Spin Lifetime Anisostropy Induced by Dynamical Symmetry Breaking}

The dynamical Rashba effect~\cite{bakulin2015real,etienne2016dynamical,glinka2020observing} offers a unique perspective for understanding the anisotropy of spin lifetime. According to the past studies~\cite{Martin2021,Niesner2018-yp,glinka2020observing,etienne2016dynamical}, the dynamical Rashba effect is induced by the local fluctuation of the organic cation MA that associatively distorts the inorganic sublattice. The local distortion within the nanometer scale at room temperature occurs on a sub-picosecond time scale, which is comparable with the spin lifetime at the same temperature. Given similar time scales of two events, the dynamical Rashba effect can inevitably affect the spin dynamics.

\begin{figure}[H]
    \centering
    \includegraphics[width=0.5\textwidth]{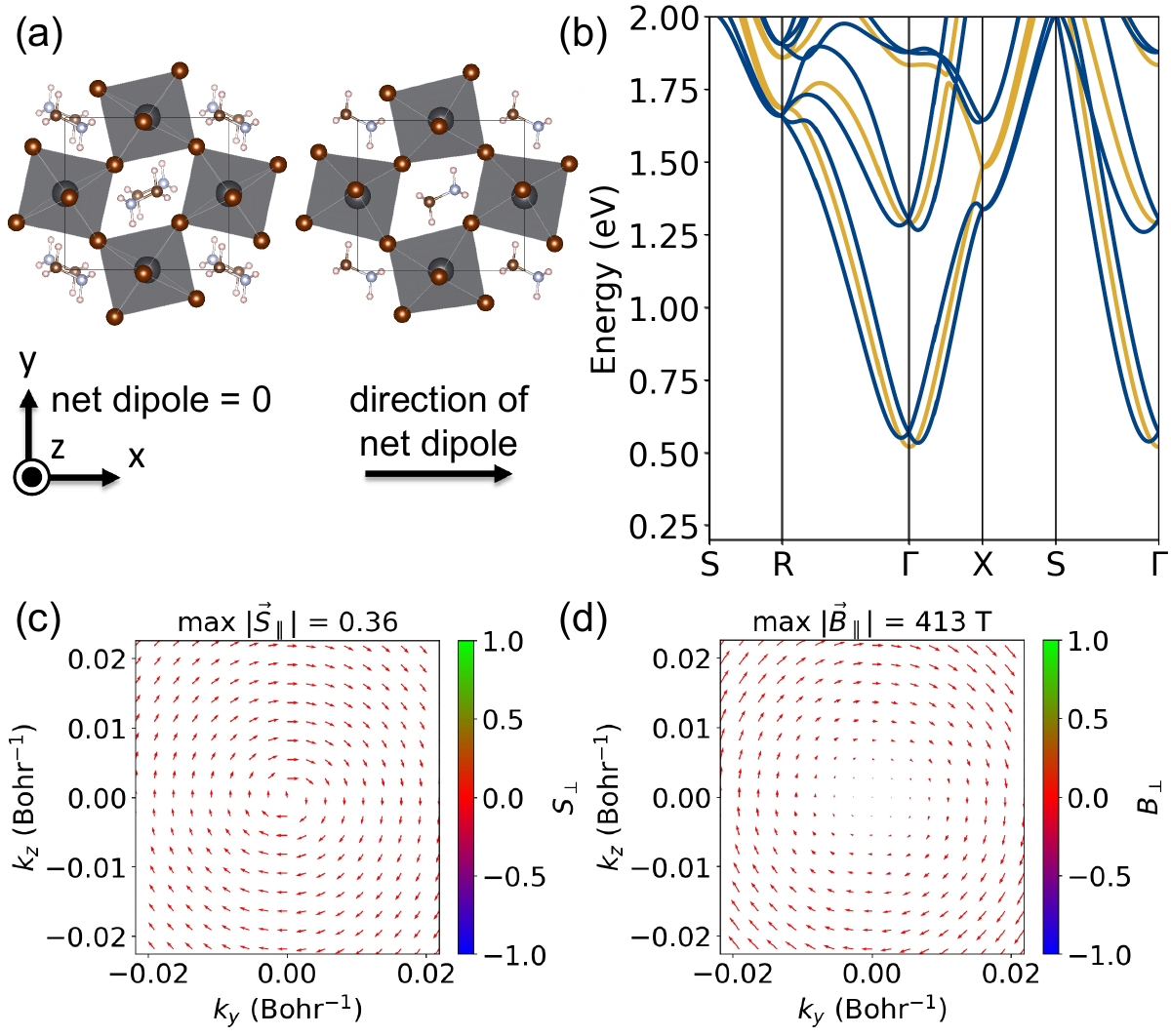}
    \caption{(a) Crystal structures of centrosymmetric (left) and asymmetric $\mapbbr$ (right), with zero and nonzero net dipole approximately pointing to x-axis, respectively. 
    (b) The band structures of the centrosymmetric $\mapbbr$ (yellow) and asymmetric $\mapbbr$ (blue). (c) The Rashba-like spin texture in the y-z plane near the conduction band minimum. (d) The internal magnetic field in the y-z plane obtained by Eq.~(\ref{eq:Bin_nk}).
    }
    \label{fig:mapbbr_crystal_structure_bandstructure_spin_texture_magnetic_field}
\end{figure}

To understand the Rashba effect on the spin dynamics, instead of iterating many possible configurations of structure, we initialized the structure of $\mapbbr$ in the maximally polarized case, where all MA molecules nearly align in x-axis then we optimized the structure, as shown in Fig.~\ref{fig:mapbbr_crystal_structure_bandstructure_spin_texture_magnetic_field}(a). Such a structure has broken inversion symmetry under structure distortion due to the MA molecule. Because of the alignment of the MA molecules, a net electric polarization occurs in nearly x-axis, which induces a large internal effective magnetic field in y-z plane. Simultaneously, as in Fig.~\ref{fig:mapbbr_crystal_structure_bandstructure_spin_texture_magnetic_field}(b), the originally degenerate Kramer's pairs split (spin split at each k point), providing a Rashba-like spin texture, as shown in Fig.~\ref{fig:mapbbr_crystal_structure_bandstructure_spin_texture_magnetic_field}(c). The following Hamiltonian $H^\text{in}$ describing the interaction between the spin and internal magnetic field (an effective Zeeman Hamiltonian) provides a way to quantify the internal magnetic field of the asymmetric $\mapbbr$,
\begin{align}
    H^\text{in} &= \frac{2g_0 \mu_B}{\hbar} \mathbf{B}^\text{in}(\mathbf{k}) \cdot \mathbf{S}_\mathbf{k} ~\label{eq:internal_magnetic_field_hamiltonian} \\
    \mathbf{B}^\text{in}(\mathbf{k}) &= C \cdot \mathbf{k}~\label{eq:C_tensor_Bin} \\
    C &= C_s + C_v + C_t.
\end{align}
Here $\mu_{B}$ is the Bohr magneton, $g_{0}$ is the free-electron $g$-factor, $\mathbf{S}_\mathbf{k}$ is the spin angular momentum, $\mathbf{B}^\text{in}(\mathbf{k})$ is the internal magnetic field vector which can be represented by the product of coefficient tensor $C$ and $\mathbf{k}$ vector. The factor of 2 accounts for the relative shift between two bands. This tensor $C$ can be decomposed into the traceless antisymmetric part (Rashba), the symmetric part (Dresselhaus at the linear order), and the trace (Weyl contributions). More details see SM Sec.~\textcolor{red}{\rn{6}}. 
As shown in Fig.~\ref{fig:mapbbr_crystal_structure_bandstructure_spin_texture_magnetic_field}(d), the internal B field is Rashba-like, dominated by the Rashba effect ($>90\%$) with a small mixture of Dresselhaus effect ($<10\%$). 
We find the Rashba splitting coefficient from fitting effective Zeeman Hamiltonian to be $1.18~\si{eV.\angstrom}$. Alternatively, we calculate the effective Rashba splitting coefficient from the band splitting by the conventional definition $\alpha=\varepsilon/(2\Delta k)=1.35~\si{eV.\angstrom}$, where $\varepsilon$ and $\Delta k$ are the energy difference between splitting bands and $k$ distance from the reference high symmetry point, respectively~\cite{etienne2016dynamical}. Note that such definition assumes the spin split is completely from the Rashba SOC contribution, which is approximately true for this system ($> 90\%$ Rashba contribution from above). 
The calculated Rashba splitting coefficients from both methods are comparable with the coefficient $1.4~\si{eV.\angstrom}$ extracted from experimental data on the same system~\cite{sajedi2020absence}.  
This provides some experimental justification for the asymmetric structural model we used here. 

\begin{figure}[H]
    \centering
    \includegraphics[width=0.4\textwidth]{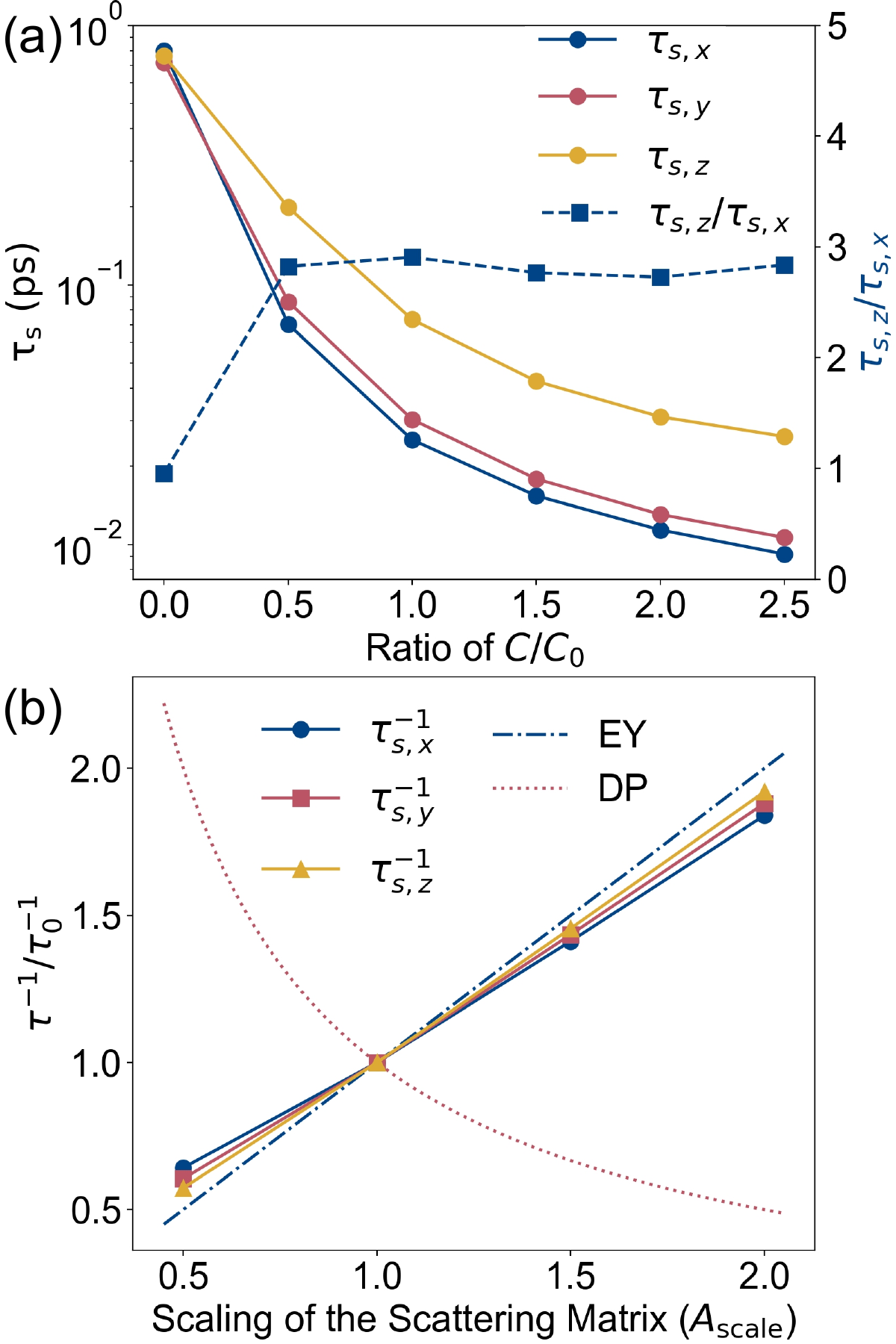}
    \caption{(a) Anisotropy of electron spin lifetime due to the internal magnetic field of asymmetric $\mapbbr$ at 300 K. The general effect that splits band structure is represented by the $C$ tensor in Eq.~(\ref{eq:C_tensor_Bin}), and $C_0$ is the fitted tensor for asymmetric $\mapbbr$. (b) Spin relaxation mechanism of asymmetric $\mapbbr$ at 300 K.}
    \label{fig:anisotropy_and_spin_relaxation_mechanism}
\end{figure}

As shown in Fig.~\ref{fig:anisotropy_and_spin_relaxation_mechanism}(a), the spin lifetime reduces monotonically by scaling up the coefficient $C$ (increasing $C/C_0$ ratio), where $C_0$ is the coefficient directly fitted from the internal magnetic field of the asymmetric $\mapbbr$.
The Rashba effect creates an inhomogeneous distribution of magnetic field in the momentum space, which is physically equivalent to the fluctuation or variation of Larmor precession ($\Delta \Omega \propto \Delta B^{in}$). This becomes another cause of spin decay ($\left(\tau_{s}^{-1}\right)^{\Delta\Omega}$) through spin precession in addition to the spin-flip relaxation ($\left(\tau_{s}^{-1}\right)|_{\Delta \langle B^\text{in}\rangle=0}$) as follows,
\begin{align}
    \tau_{s}^{-1} \approx \left(\tau_{s}^{-1}\right)|_{\Delta \langle B^\text{in}\rangle=0} + \left(\tau_{s}^{-1}\right)^{\Delta\Omega} \label{eq:spin_relaxation_spin_flip_spin_precession}
\end{align} 

In particular, the Rashba-like SOC leads to the anisotropy of spin lifetime, i.e., the spin lifetime along y or z direction (in-plane) is longer than that in x direction (out-of-plane). The reason is that the fluctuation of internal magnetic field ($\Delta \langle B^\text{in}\rangle$) is stronger in the y-z plane than in the x-y and x-z planes, i.e. $\Delta  \langle B^\text{in}\rangle_{xy}=57.3$ T, $\Delta \langle B^\text{in}\rangle_{xz}=119.1$ T and $\Delta \langle B^\text{in}\rangle_{yz}=131.5$ T. Here the internal magnetic field $\bf{B}^{in}$ is evaluated according to
\begin{align}
    \mathbf{B}^\text{in}_{nk} &= \frac{2\Delta_{nk} \mathbf{S}^\text{exp}}{g_e\mu_B} \label{eq:Bin_nk}\\
    \langle B^\text{in}\rangle_i &= \sum_{nk} \frac{f'_{nk} B^\text{in}_{i;nk}}{\sum_{nk} f'_{nk}} \label{eq:Bin}\\
    \Delta \langle B^\text{in}\rangle_i &= \sqrt{\sum_{nk} \frac{f'_{nk} (B^\text{in}_{i;nk} - \langle B^\text{in}\rangle_i)^2}{\sum_{nk} f'_{nk}}} \label{eq:delta_Bin}\\
    \Delta \langle B^\text{in}\rangle_{ij} &= \sqrt{\Delta \langle B^\text{in}\rangle_{i}^2 + \Delta \langle B^\text{in}\rangle_{j}^2}
\end{align}
%
%
where $\mathbf{B}^\text{in}_{nk}$ is the internal magnetic field as the product of the band splitting $\Delta_{nk}$ and spin expectation value $\mathbf{S}^\text{exp}$ at band $n$ and $k$ point. 
$\langle B^\text{in}\rangle_i$ and $\Delta \langle B^\text{in}\rangle_i$ are the thermal average of the internal magnetic field and magnetic field fluctuation in $i$-th Cartesian direction, respectively, with $f'$ being the derivative of Fermi-Dirac distribution function~\cite{xu2021giant}.

As the  magnetic field fluctuation orthogonal to the spin direction of interest leads to spin relaxation or dephasing, stronger $\Delta \langle B^\text{in}\rangle$ effectively gives rise to faster Larmor precession and shorter spin lifetime. The weakest $\Delta \langle B^\text{in}\rangle_{xy}$ accounts for the longest spin lifetime $\tau_{s,z}$. The stronger and similar $\Delta \langle B^\text{in}\rangle_{xz}$ and $\Delta \langle B^\text{in}\rangle_{yz}$ account for the similarity between the shorter $\tau_{s,x}$ and $\tau_{s,y}$. 
Interestingly, the spin lifetime anisotropy aligns with the experimental observation~\cite{huynh2024magneto} of ${\sim}2.5$ across a wide range of the internal B field coefficient, as seen in Fig.~\ref{fig:anisotropy_and_spin_relaxation_mechanism}(a). This suggests that the spin lifetime anisotropy has weak dependence on the size of Rashba coefficients, generated by the lattice polarization from molecular rotation.

\subsection{Strong Spin Lifetime Anisotropy from Persistent Spin Helix}

\begin{figure}
    \centering
    \includegraphics[width=0.5\textwidth]{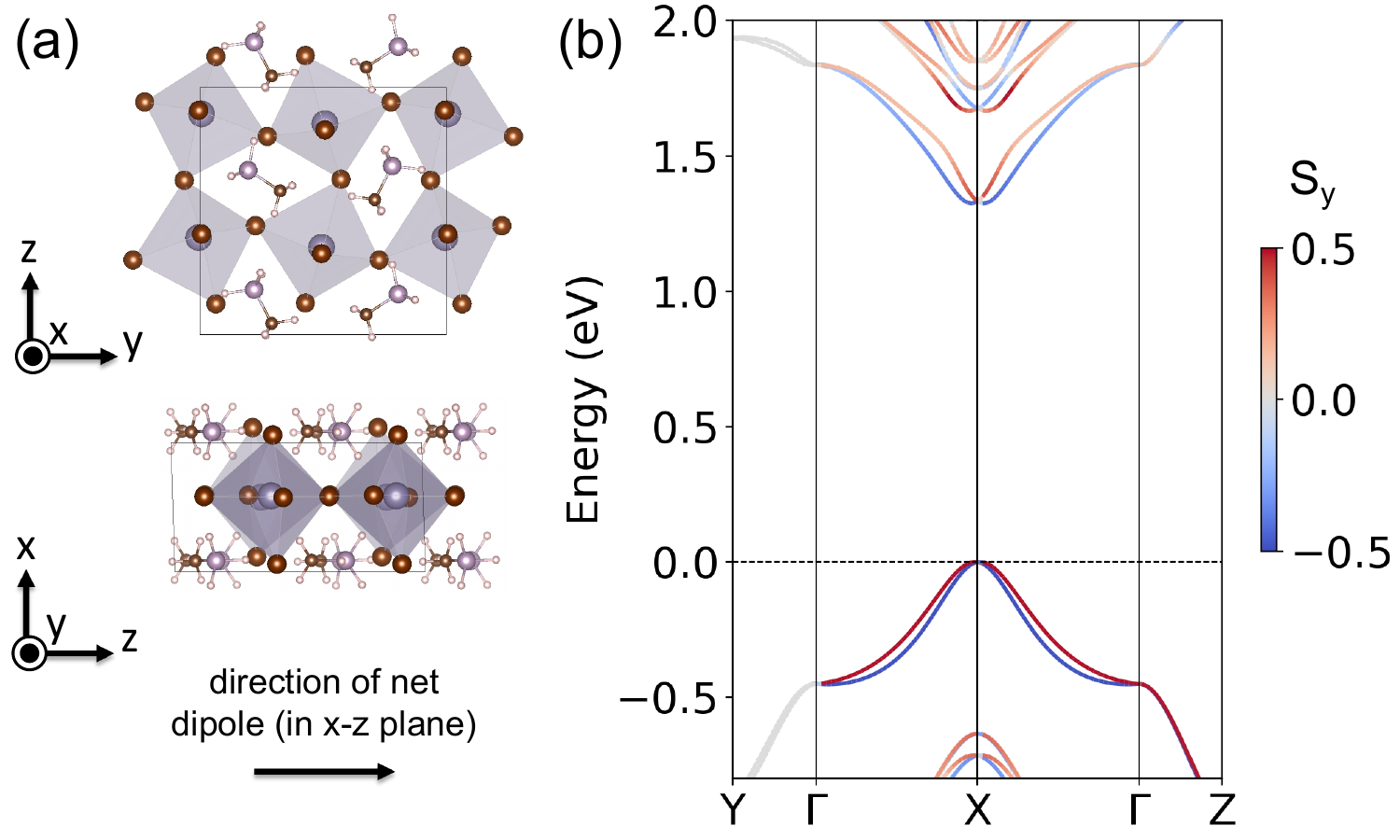}
    \caption{(a) Crystal Structure of $\mpsnbr$ of $P_c$ crystal symmetry. The net dipole due to the alignment of the MP molecules is in the x-z plane and mostly along the z-axis~\cite{kashikar2022rashba}. (b) Band structure of $\mpsnbr$, showing spin polarization projected along the y-axis. The S$_x$ and S$_y$ spin polarization projection on band structure can be found in SM Fig.~\textcolor{red}{S9}.   
    }
    \label{fig:mpsnbr_structure_bandstructure}
\end{figure}

\begin{figure}[H]
    \centering
    \includegraphics[width=0.4\textwidth]{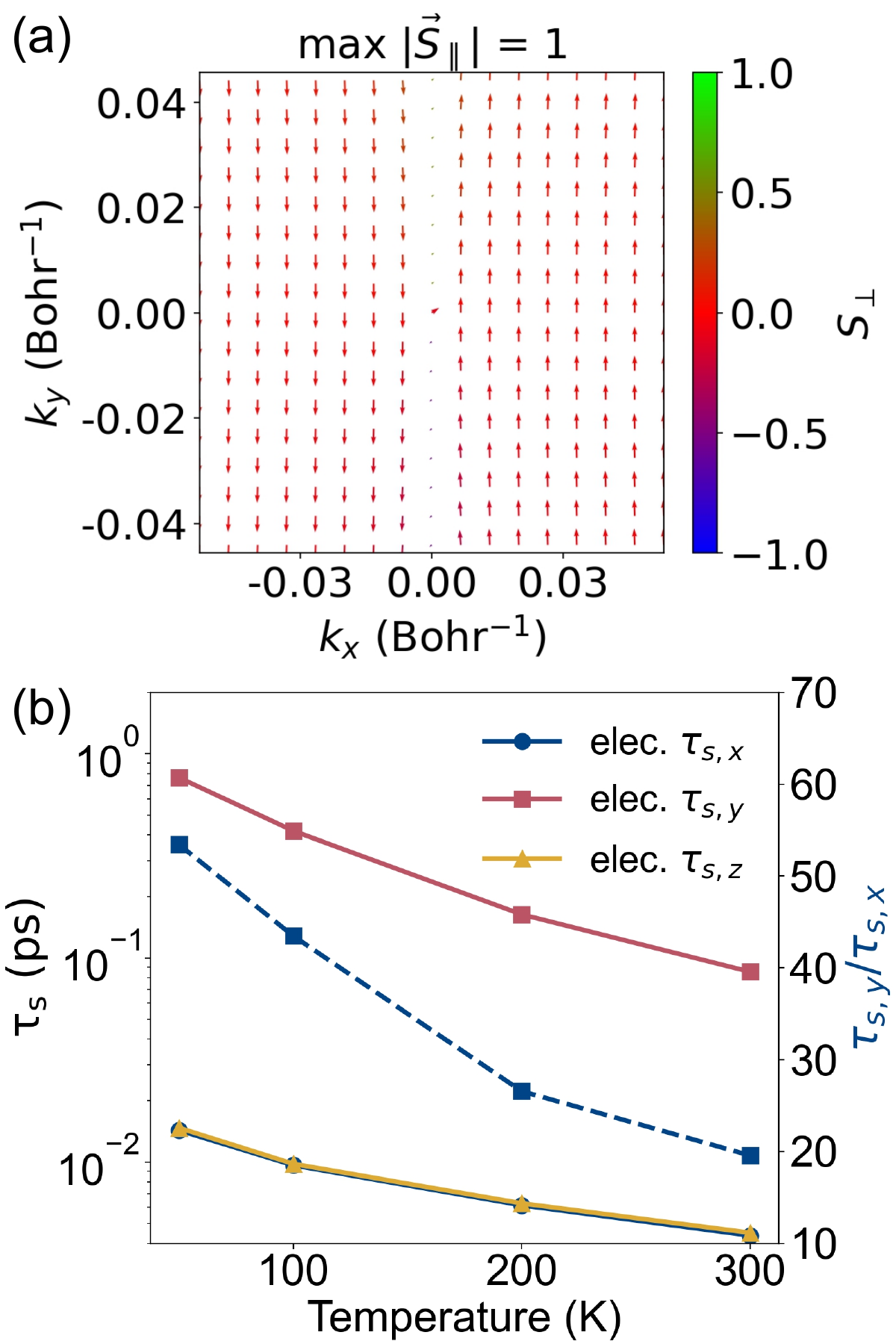}
    \caption{
    (a) The PSH spin texture at CBM  centered at high symmetry point X. (b) Temperature-dependent spin lifetime of electrons in $\mpsnbr$ at zero external magnetic field. The red, blue and yellow data points with solid lines represent the spin lifetime with polarization along x, y and z directions. The blue data points with the dashed line are the anisotropy of spin lifetime with polarization between y (parallel to PSH) and x (perpendicular to PSH) directions. 
    }
    \label{fig:mpsnbr_spin_lifetime_spin_texture}
\end{figure}

%
Next, we will show that the PSH spin texture can give rise to a significantly larger anisotropy in spin lifetime, using an example of $\mpsnbr$, which has a crystal symmetry of $Pc$ and a corresponding point group of $C_s$, as shown in Fig.~\ref{fig:mpsnbr_structure_bandstructure}(a). Compared to MA molecules, the MP molecule is more polarized and less likely to rotate around its center due to the high rotation energy barrier~\cite{zhang2020methylphosphonium}. 
The highly polarized and stable alignment of the MP molecules give rise to a strong net dipole in the x-z plane and mostly along z-axis, which subsequently results in strong distortion in the inorganic sublattice and a permanent symmetry breaking. This results in an internal magnetic field in the x-y plane. Due to the $C_s$ symmetry, a significant result is the persistent spin helix in the conduction band~\cite{kashikar2021chemically,kashikar2022rashba,lu2023strain}, with spin highly polarized in the y-axis, as shown in Fig.~\ref{fig:mpsnbr_structure_bandstructure}(b) and Fig.~\ref{fig:mpsnbr_spin_lifetime_spin_texture}(a). 
We numerically fitted the internal B field Hamiltonian in Eq.~\ref{eq:C_tensor_Bin} to confirm PSH, where Rashba and Dresselhaus contributions are nearly equal. Details can be found in SM Sec.~\textcolor{red}{X}. 


As shown in Fig.~\ref{fig:mpsnbr_spin_lifetime_spin_texture}(b), the spin lifetime of electrons along the y direction is 20-50 times longer than the spin lifetime along the x and z directions. 
This remarkable spin lifetime anisotropy can be explained by the suppression of spin scattering in the direction of PSH~\cite{ohno1999spin,bernevig2006exact}. Additionally, the spin lifetime of electrons in $\mpsnbr$ is generally one order of magnitude shorter compared to that of centrosymmetric $\mapbbr$. This can be attributed to the exceptionally large internal magnetic field (maximally 956 T in the momentum space) of $\mpsnbr$ and the small range of PSH in momentum space (approximately 20\% of the first Brillouin zone, SM Fig.~\textcolor{red}{S10}), 
which quickly transitions to a Rashba-like spin texture as it moves away from the band edge~\cite{kashikar2022rashba}. Consequently, there occurs the fluctuation of internal magnetic field, and $\Delta  \langle B^\text{in}\rangle_{xz}=38.1$ T is one order smaller than $\Delta  \langle B^\text{in}\rangle_{xy}=508.5$ T and $\Delta  \langle B^\text{in}\rangle_{yz}=509.5$ T. The fluctuation of the internal magnetic field leads to additional spin relaxation or dephasing that reduce spin lifetime, as described by Eq.~(\ref{eq:spin_relaxation_spin_flip_spin_precession}), and longer spin lifetime along the y direction than z and x.

In contrast, as can be seen in SM Fig.~\textcolor{red}{S13}(a), the spin lifetime of holes at VBM is less anisotropic, and similar to the spin lifetime anisotropy of asymmetric 
$\mapbbr$. This anisotropy is attributed to the Rashba-like spin texture at VBM in SM Fig.~\textcolor{red}{S13}(b). 
%
%
%
%
Compared to the spin lifetime of holes, the spin lifetime of electrons in the y direction is more intringing, highlighting the fact that the spin lifetime can be more reliably controlled by the PSH. The attenuated protection of the PSH to spin relaxation suggests the need of materials that exhibit a larger range of PSH in the $k$ space.

\section{Conclusion}
In summary, we present the spin dynamics study of hybrid inorganic-organic perovskites, $\mapbbr$ and $\mpsnbr$, with a particular focus on elucidating the anisotropy of spin lifetime with the dynamical Rashba effect and permanent symmetry breaking.

We first present the theoretical spin lifetime of the centrosymmetric $\mapbbr$, which agrees with experimental observations, particularly in terms of temperature and magnetic field dependence. The EY, DP, and FID mechanisms are identified as the primary origins of spin relaxation and dephasing under zero, weak, and strong external magnetic fields, respectively. Besides, we reveal the distinct phonon mode dependence in carrier and spin relaxation. 

However, the anisotropy of spin lifetime in $\mapbbr$ is much weaker than what is observed experimentally, suggesting additional mechanism yet to be identified. 
Specifically, we attribute the observed anisotropy to the dynamical Rashba effect due to molecular rotation, which occurs on a similar timescale as the spin lifetime, and has been observed in several past experiments. Our structural model for studying dynamical Rashba effect is supported by the consistent Rashba coefficient between theoretical and experimental values. The symmetry breaking induced by the Rashba effect introduces additional DP/FID spin dephasing alongside the EY mechanism, resulting in the observed anisotropy in spin lifetime.

Finally, we calculate the spin lifetime of electrons and holes in $\mpsnbr$ and find a large anisotropy of spin lifetime of electrons, as a result of permanent symmetry breaking. A significant enhancement of spin lifetime is observed in the parallel direction to PSH compared to the perpendicular directions. Even though the spin lifetime of $\mpsnbr$ is shorter than the widely studied $\cspbbr$ and $\mapbbr$, our theoretical results provide in-depth understanding of the PSH effect on spin lifetime, and insights into the design of  spintronics materials. We propose materials with strong SOC and a wide-range of PSH in k space to enable the simultaneous effective spin control and long spin lifetime for optimal spintronics applications.


\section{Computational Methods}
We perform electronic structure, phonon, Wannier calculations by using Quantum-Espresso~\cite{QE} and JDFTx~\cite{sundararaman2017jdftx}.
We choose the energy cutoff of wavefunction 74 Ry using the PseudoDojo Optimized Norm-Conserving Vanderbilt (ONCV) pseudopotential~\cite{hamann2013optimized,van2018pseudodojo}. We use the exchange-correlation functionals with Perdew-Burke-Ernzerhof (PBE) generalized gradient approximation~\cite{PBE1997} to optimize the lattice constants and internal geometry, and obtain phonon properties as well as single-particle electronic structure. 
We calculate the phonon and electron-phonon interaction by using the finite difference method with self-consistent SOC from the fully relativistic pseudopotentials~\cite{xu2021ab}. 
The perovskites involve the orthorhombic $\mapbbr$ of the $Pnma$ space group~\cite{swainson2003phase}, and the monoclinic $\mpsnbr$ of the $Pc$ space group~\cite{zhang2020methylphosphonium}.

We simulate spin and carrier dynamics using the method of first-principles density-matrix dynamics (FPDM)~\cite{xu2020spin,xu2021ab}. This involves solving a quantum master equation that governs the time evolution of the density matrix $\rho(t)$, depicted as follows:
\begin{align}
\begin{split}
\frac{d\rho_{12}(t)}{dt} &= \big[H(\mathrm{{\mathbf{B}}}),\rho(t)\big]_{12} \\
&\quad+ \frac{1}{2} \sum_{345} \Big\{ \big[I-\rho(t)\big]_{13}P_{32,45}\rho_{45}(t) \\
&\quad- \big[I-\rho\left(t\right)\big]_{45}P_{45,13}^{*}\rho_{32}(t) \Big\} + H.C.
\end{split}~\label{eq:master}
\end{align}
Here the equation incorporates Larmor precession in the first term, and scattering processes in the second term. Within the first term, $H(\mathbf{B})$ represents the electronic Hamiltonian in the presence of a magnetic field $\mathbf{B}$, and $[H,\rho]=H\rho - \rho H$. The second  term includes the general mechanisms that induce spin relaxation through the spin-orbit coupling (SOC). The generalized scattering-rate matrix $P$ encompasses electron-phonon, electron-impurity, and electron- electron scattering processes including spin-orbit couplings, computed from first-principles~\cite{Xu2024JCTC}. The numeric indices are the contracted notation for the k-point and band of electronic states. The notation of $H.C.$ signifies Hermitian conjugate. Beginning with an initial density matrix initialized with a net spin, we evolve $\rho(t)$ over sufficient duration, typically ranging from hundreds of picoseconds to a few microseconds, using the Eq.~\ref{eq:master}. Subsequently, we compute the spin observable $S(t)$ from $\rho(t)$ and determine the spin lifetime from fitting $S(t)$. More theoretical background and details can be found in Ref.~\cite{Xu2024JCTC}.

\section{Experimental Methods}
The $\mapbbr$ thin films were prepared by using a precursor solution. Precursors $\mathrm{CH_3NH_3Br}$ and $\mathrm{PbBr_2}$ are mixed in a molar ratio of 1:1 in solvant N,N-dimethylformamide, and the concentration of 0.8 mol/ml~\cite{huynh2024magneto}. The $\mapbbr$ crystals were grown via antisolvent precipitation. X-ray-diffraction (XRD) measurements confirm the crystal structure and principal axes.

The ultrafast circularly-polarized photoinduced reflectivity (c-PPR) method~\cite{HuynhPRB2022,huynh2024magneto} is used to measure the spin lifetime of carriers in $\mapbbr$ single crystals. The PPR method is a derivative of the pump-probe technique, where the polarization of the pump beam is modulated by a photoelastic modulator between left circular polarization (LCP, $\delta^{+}$) and right circular polarization (RCP, $\delta^{-}$), and the probe beam is modulated to LCP/RCP by a quarter-wave plate. The pump beam is of 405 nm wavelength, 250 femtoseconds pulse duration, and 80 MHZ repetition rate. It is generated by frequency doubling the fundamental at 810 nm from the Ti:Sapphire laser (Spectra Physics) using a second-harmonic generator (SHG) barium borate (BBO) crystal. The 533~nm probe beam was generated by combining the 810 nm fundamental beam with the 1560 nm infrared beam from an optical parametric amplifier (OPA) onto a BBO type 2 SFG (Sum Frequency Generation) crystal. Using this technique we measured both t-PPR responses at both zero and finite $B$ to extract the $B$-dependent spin lifetimes of electrons.

\begin{figure}
{
    \centering
    \includegraphics[width=0.5\textwidth]{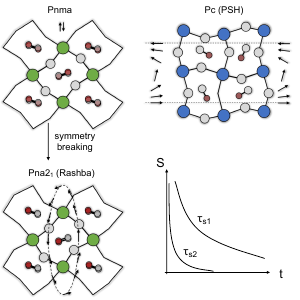}
\captionsetup{justification=justified,singlelinecheck=false}
    \caption{Table of Content: Strong spin lifetime anisotropy with broken inversion symmetry structures. }
    }
    \label{fig:TOC}
\end{figure}

\begin{acknowledgement}
We acknowledge the support for materials theory application and experimental measurements of spin dynamics as part of the Center for Hybrid Organic-Inorganic Semiconductors for Energy (CHOISE), an Energy Frontier Research Center funded by the Office of Basic Energy Sciences, Office of Science within the US Department of Energy (DOE). We acknowledge support for the theoretical development of spin dynamics in the presence of large spin-orbit coupling and magnetic field by the computational chemical science program within the Office of Science at DOE under grant No. DE-SC0023301. 

This research used resources of the Center for Functional Nanomaterials, which is a US DOE Office of Science Facility, and the Scientific Data and Computing center, a component of the Computational Science Initiative, at Brookhaven National Laboratory under Contract No. DE-SC0012704, the lux supercomputer at UC Santa Cruz, funded by NSF MRI grant AST 1828315, the National Energy Research Scientific Computing Center (NERSC) a U.S. Department of Energy Office of Science User Facility operated under Contract No. DE-AC02-05CH11231, and the Extreme Science and Engineering Discovery Environment (XSEDE) which is supported by National Science Foundation Grant No. ACI-1548562 \cite{xsede}.
\end{acknowledgement}

\bibliography{ref}

\end{document}